\documentclass{iopjournal}
\usepackage{xspace}
\usepackage{graphicx}
\usepackage{hyperref}
\usepackage{xurl} 
\usepackage[T1]{fontenc}
\usepackage{lmodern}
\usepackage{amsmath} 
\usepackage{amssymb} 
\usepackage[numbers,sort&compress]{natbib}
\usepackage{xcolor}
\usepackage[normalem]{ulem}
\usepackage{soul}
\setstcolor{red}

\newcommand{\bbh}{\texttt{BBH}\xspace}
\newcommand{\bbhs}{\texttt{BBHs}\xspace}

\newcommand{\hu}{\ensuremath{{\rm km s^{-1} Mpc^{-1}}}\xspace}

\newcommand{\bpop}{\textsc{B-pop}~}

\begin{document}

\articletype{Article} 

\title{Calibrating spectral siren cosmology with synthetic catalogs of binary black hole mergers}

\author{Arianna Scarpa$^{1}$\orcid{0009-0001-0346-7920} , Simone Mastrogiovanni$^1$\orcid{0000-0003-1606-4183}, Filippo Santoliquido$^{2,3}$\orcid{0000-0003-3752-1400}, Manuel Arca Sedda$^{2,3,4}$\orcid{0000-0002-3987-0519}},  Giovanni Antinozzi$^{5,6}$\orcid{0009-0005-3537-682X} \\ 
\affil{$^1$INFN, Sezione di Roma, 1-00185 Roma, Italy}
\affil{$^2$Gran Sasso Science Institute, Via F. Crispi 7, I-67100 L’Aquila, Italy}
\affil{$^3$INFN, Laboratori Nazionali del Gran Sasso, I–67100 Assergi, Italy}
\affil{$^4$INAF, Osservatorio Astronomico di Roma, I-00040 Monte Porzio Catone (Rome), Italy}
\affil{$^5$SISSA, via Bonomea 265, 34136 Trieste, Italy}
\affil{$^6$INFN, Sezione di Trieste, Via Valerio 2, 34127 Trieste, Italy}\\

\affil{$^*$Author to whom any correspondence should be addressed.}

\email{arianna.scarpa@ligo.org}


\begin{abstract}
Binary black hole (\bbh) mergers detected through Gravitational Waves (GWs) are a promising probe for the cosmic expansion. These sources are standard sirens for which we can directly measure the luminosity distance, but their redshift is degenerate with the determination of their source masses. In analogy to standard candles, the redshift of standard sirens can be obtained using a calibration based on the source mass spectrum, but without the need for a cosmological ladder. 
The modelling of the \bbh mass spectrum is crucial for GW cosmology. In this paper, we propose a new pipeline based on Normalizing Flows to calibrate population models on synthetic \bbh catalogs generated from astrophysical prescriptions. Although these models are less flexible than the current parametric models used in the literature, they have the advantage of directly capturing the correlations between \bbh mass and redshift and provide a direct link between different formation channels and cosmological parameter inference. We discuss potential applications and limitations of this method using the \bpop code to generate a mock catalog that contains \bbh formed by isolated stellar binaries and in dense stellar environments via dynamical interactions.
Using this methodology, we show that \bpop provides a good fit for the mass and redshift distributions of the GW events in GWTC-4.0 data for a flat $\Lambda$CDM cosmology with $H_0 \sim 70 \hu$ and $\Omega_m=0.308$. We also find a correlation between $H_0$ and the fraction of binaries produced by the dynamical and isolated formation channels arising from the fact that dynamically formed binaries populate the higher side of the mass spectrum.
\end{abstract}

\section{Introduction}

Gravitational Wave (GW) sources are rapidly becoming promising probes of the cosmic expansion. This is particularly relevant given the unresolved puzzle in modern cosmology known as the ``Hubble tension'' \cite{hubbletension}: measurements of the Hubble constant $H_0$ differ between direct measurements in the local Universe and indirect estimates based on early universe observations from the Cosmic Microwave Background \cite{plank}.

GWs from compact binary coalescences (CBCs) provide a direct measurement of the luminosity distance to the source, without the need for a cosmological ladder \cite{firstH0cosmo, firstgwstandardsirens}, unlike standard candles such as supernovae \cite{shoes}. However, they do not provide a direct estimation of the redshift that could be obtained from the identification of the source host galaxy, as in the case of GW170817 \cite{LIGOScientific:2018gmd, natureH0GW}. When no electromagnetic counterpart is detected, the source redshift can be inferred statistically, either by using the source frame mass properties - also known as spectral siren method -- of binary black holes (\bbhs) \cite{Taylor:2012db, You:2020wju, Mancarella:2021ecn, Ye:2021klk,  Ezquiaga:2022zkx, Mastrogiovanni:2023emh, Dalang:2024gfk, Mali:2024wpq, MaganaHernandez:2025cnu}, galaxy catalogs \cite{Palmese:2020}, or cross-correlation \cite{Schutz:1986gp, firstgwstandardsirens, Gray:2023wgj, Oguri:2016dgk, Mukherjee:2020hyn, Bera:2020,Mukherjee:2022afz} with large-scale structure tracers.

With the release of the fourth Gravitational Wave Transient Catalog (GWTC-4.0) \cite{gwtc4}, the LIGO-Virgo-KAGRA (LVK) collaboration has confirmed that the mass distribution of \bbhs is one of the most promising avenues to measure the cosmic expansion \cite{gwtc4cosmo}. The mass spectrum of \bbhs is far from trivial \cite{gwtc4pop} to describe; it can contain underdensity and overdensity of \bbh production. These features might arise from the astrophysical formation mechanisms of \bbhs.
Population studies are becoming increasingly precise thanks to the rapidly growing number of detections. Despite these advances, population modeling remains a major challenge, as modeling compact binary formation channels is intrinsically complex. Recent works have highlighted tensions in the literature regarding correlations between binary parameters such as mass, spin, and redshift \cite{evidence_red_ev_1, evidence_red_ev_2, evidence_red_ev_3, evidence_red_ev_4, spincorr}, and this is because \bbh populations are expected to belong to distinct astrophysical formation channels and to change at different cosmic times because of metallicity and stellar evolution processes \cite{mapellibbhformationchannels, astro_red_ev_pop_1}. 
The current GW inference on cosmological parameters makes use of phenomenological prescriptions \cite{gwtc4pop} to describe the \bbh population. This approach is currently named ``strongly modelled''. Analytical phenomenological models such as Power Laws and Gaussian Peaks could not capture the complex behaviour of the \bbh mass spectrum. To alleviate this issue, a possible solution is to employ non-parametric or semi-analytical models for the mass distribution \cite{MaganaHernandez:2024uty,Farah:2024xub, MaganaHernandez:2025cnu, Tagliazucchi:2026gxn}. However, these models are not directly linked to astrophysical formation mechanisms, and if they do not capture the redshift evolution of the mass distribution, this mismodeling is shown to introduce systematic biases in the inferred value of the Hubble constant $H_0$ \cite{Mukherjee:2021rtw, Karathanasis:2022rtr, Agarwal:2024hld, pierrasystematics}. 

This work directly addresses this issue by introducing a population model calibrated on state-of-the-art population synthesis simulations of \bbhs. Catalogs of \bbhs naturally include all the correlations between masses, redshift, and other binary parameters that could arise from astrophysical formation channels. Our purpose is to connect simulations of \bbh, and their respective formation channels to cosmological inference. This type of approach complements the ones in current literature, where parametric \cite{gwtc4cosmo, gwtc4pop}, or non-parametric population models \cite{2025ApJ...978..153F, 2026A&A...709A.197T, 2026arXiv260306792B} are used for cosmological inference.

The models that we consider are built using Normalizing Flows. These are powerful machine learning neural networks capable of learning complex probability distributions directly from data, without imposing analytical functional forms. This flexibility is required calibrating population models on complex astrophysical simulations of \bbh to GW observations. More generally, Normalizing Flows have been employed in recent works to emulate astrophysical simulations and to perform population inference from GW catalogs \cite{NF_GW_1, NF_GW_2, NF_GW_3, NF_GW_4, NF_GW_5}, without considering the cosmological parameters.

This paper is organized as follows. In Sec~\ref{HBI e NF} we introduce the hierarchical Bayesian inference framework, the spectral sirens method, the non-parametric population model based on Normalizing Flows, and we validate the model using a mock catalog with a redshift-evolving population. In Sec~\ref{bpop} we apply the model to an astrophysical synthesis catalog and to simulated and real gravitational wave observations, jointly constraining cosmological and population parameters. Our findings are finally summarized in Sec~\ref{conclusions}.

\section{Overview of the methodology and the pipeline} \label{HBI e NF}

In this section, we provide an overview of the pipeline that we designed to connect mock catalogs of \bbh 
binaries to cosmological inference.

\subsection{Hierarchical Bayesian inference and Normalizing Flows}

Given a set of observed gravitational wave data $\{x_i\}$, the likelihood for the population parameters $\Lambda$ is given by the so-called \textit{hierarchical likelihood} \cite{icarogw}:
\begin{equation}
    \mathcal{L} (\{x\}|\Lambda) \propto e^{-N_{\rm exp}(\Lambda)} 
    \prod_{i=1}^{N_{\rm obs}} 
    \int \mathcal{L}(x_i|\theta, \Lambda) 
    \frac{\mathrm{d} N}{\mathrm{d} \theta \mathrm{d} t}(\Lambda) \mathrm{d} \theta \mathrm{d} t ,
    \label{hierarchical likelihood}
\end{equation}
where $\theta$ represents the binary parameters, $\mathcal{L}(x_i| \theta,\Lambda)$ is the single-event likelihood, and $T_{\rm obs}$ is the observation time. $N_{\rm exp}(\Lambda)$ represents the expected number of detected events and accounts for selection effects. It is defined as:
\begin{equation}
    N_{\rm exp}(\Lambda) = T_{\rm obs} 
    \int \frac{\mathrm{d} N}{\mathrm{d} \theta \mathrm{d} t}(\Lambda) \; \mathrm{d} \theta  
    \int_{x \in \rm detectable} 
    \mathcal{L}(x_i|\theta, \Lambda) \mathrm{d} x .
    \label{nexp}
\end{equation}
The integral over detectable events represents the probability that a signal with parameters $\theta$ is detected, and it accounts for the selection bias \cite{selectionbias} caused by the finite sensitivity of gravitational wave detectors \cite{hbi1, hbi2, gwcosmologyintroduction}.

The population model enters the hierarchical likelihood through the detector-frame rate $\frac{dN}{d\theta dt}$, which depends on cosmological parameters and on the population parameters that describe the intrinsic properties of the sources. This method is the so-called \textit{spectral sirens} analysis \cite{spectralsirens}, in which cosmological information is extracted from gravitational wave observations by exploiting the relation between the detector-frame and source-frame parameters, without the need for an electromagnetic redshift measurement. The detector-frame rate is usually factorized as:
\begin{equation}
    \frac{\mathrm{d} N}{\mathrm{d} d_L \mathrm{d} m_{1d} \mathrm{d} m_{2d} \mathrm{d} t_d} =
    R_0 \; p_{\rm pop}(m_{1s}, m_{2s}|\Lambda, z) 
    \Psi(z, m_{1s}, m_{2s} | \Lambda)
    \frac{\mathrm{d} V_c}{\mathrm{d} z} 
    \frac{1}{1+z} 
    \frac{1}{|J_{d\rightarrow s}|} ,
    \label{cbcrate}
\end{equation}
where $R_0$ is the local merger rate density, $p_{\rm pop}(m_{1s}, m_{2s}|\Lambda, z)$ and $\Psi(z, m_{1s}, m_{2s}|\Lambda)$ are the distributions describing the joint primary and secondary source-frame mass distributions and the redshift evolution of the merger rate, respectively. The factor $dV_c/dz$ is the differential comoving volume element, while the term $1/(1+z)$ and the Jacobian $|J_{d\rightarrow s}|$ account for the conversion from detector-frame to source-frame parameters. 
In most GW population analyses, $p_{\rm pop}$ is assumed to be redshift-independent and $\Psi$ is assumed to be mass-independent,  and are modelled with simple parametrized forms \cite{gwtc4pop, gwtc3pop}. When such mass/redshift correlations are not taken into account, the mismatch between the true population and the assumed phenomenological model can lead to systematic biases in hierarchical inference \cite{Mukherjee:2021rtw, Karathanasis:2022rtr, Agarwal:2024hld, pierrasystematics}.
These correlations are non-trivial to model as they arise from the several formation channels of \bbh binaries.

In this work, we model $p_{\rm pop}$ and $\Psi$ with Normalizing Flows (NFs): a machine learning technique which allows us to describe the joint distribution of \bbh source-frame masses and merger redshift from mock catalogs without using phenomenological prescriptions. NFs are generative machine learning models \cite{nffirst, nfintro, nfinference} that allow us to compute complex probability density functions through invertible and differentiable transformations between a latent distribution and the observable parameter space. In this work, the latent variable $u \in \mathbb{R}^3$ is drawn from a multivariate standard Gaussian distribution $p_U(u)=\mathcal{N}(0,\mathbb{I})$ and mapped to the observable variables $x=(m_{1s}, m_{2s}, z)$ through an invertible transformation $x=f(u)$. The resulting probability density is obtained with the change-of-variables formula:
\begin{equation}
    p_X(x) = p_U\big(f^{-1}(x)\big)\,
    \big|\det J_{f^{-1}}(x)\big|
    = p_U(u)\,\big|\det J_f(u)\big|^{-1},
\end{equation}
where $J_{f^{-1}}$ is the Jacobian of the inverse transformation. The transformation $f$ is built with the \textsc{zuko} library, using Neural Spline Flows (NSF) \cite{nsf}, in which the mapping is parametrized by autoregressive, monotonic rational-quadratic spline transformations \cite{maf}. More details about the NF's architecture, renormalization, and training are given in App.~\ref{appendix nf architecture and training}. 

The NF is used to build a population model that can be defined as
\begin{equation}
p_{\rm pop}(m_{1s}, m_{2s}|\Lambda, z) \Psi(z, m_{1s}, m_{2s} | \Lambda)
\;=\; k \;p_{\rm NF}(m_{1s},m_{2s},z),
\label{eq:nfmod}
\end{equation}
where $p_{\rm NF}(m_{1s},m_{2s},z)$ is the joint probability density function computed with the NF and $k$ is an overall normalization factor used to calculate the physical merger rate.
This formulation directly models correlations between \bbh source-frame masses and merger redshift. The model is trained on astrophysical \bbh synthesis catalogs to consistently incorporate redshift-dependent mass evolution within the HBI framework.
This approach can also be extended in a scenario where we are provided with catalogs reporting different formation channels. In this case, we can define a mixture fraction $\phi$ \cite{mastrogiovannistellarprogenitors} such that
\begin{equation}
    p_{\rm NF}(m_1,m_2,z) = \phi \; p_{\rm NF}^{\rm A}(m_1,m_2,z) + (1-\phi)\; p_{\rm NF}^{\rm B}(m_1,m_2,z),
    \label{eq:mix}
\end{equation}
where $\rm A,B$ indicate two separate formation channels. The mixture fraction is a population parameter of the hierarchical likelihood and can be used to understand what type of formation channel contributes the most to a particular value of the cosmological parameters. In principle, one could also parametrize $\phi(m_1,m_2,z)$ as a function of masses and redshift; however, this choice would correspond to the case in which we want to add some corrections to the simulated formation channels. We do not consider this case as it would introduce the requirement of additional modelling for $\phi$.

\subsection{The pipeline}

Before considering the NF's population models for application on real data, we should ensure that the fitted model is accurate enough for spectral siren cosmology. This is achieved with an iterative procedure which is used to optimize and calibrate the NF's architecture.

First, a set of architectures with varying number of transformation layers, hidden neurons, learning rate, and batch size is chosen. For each configuration, the NF is trained on a catalog by minimizing negative log likelihood (NLL) on the training set, and the corresponding validation loss is evaluated. The training procedure is described in App.~\ref{appendix nf architecture and training}. Following the training of the NFs, we validate them on mock data generated from the training set. The mock data consists of a set of observed \bbh binaries in a flat $\Lambda$CDM universe with a fixed value $H_0$ and $\phi$ set to the nominal value of the catalog; see Sec.~\ref{calibration of the NF} for more details. The final model is selected as the one that minimizes the NLL and recovers the injected $H_0$ and $\phi$ within a 90\% C.I. credible interval.

This procedure is necessary as we noticed that different architectures with NLL differing up to 5\% can perform very differently for spectral siren analyses. Models with the lowest NLL fit the overall distribution better, but they do not always recover the injected values of $H_0$ and $\phi$ that are used when we generate mock GW observations from the catalogs. This is because small discrepancies in the low-redshift tails, that is, where GW observations lie, can affect the recovery of the injected values. Whenever possible, the lightest architecture is chosen to minimize computational times.

The entire pipeline procedure is demonstrated in a case study reported in App.~\ref{appendix mock catalog benchmark} where we considered a redshift-evolving mock \bbh source population, and then simulate GW observations from it and infer $H_0$.
We considered two different scenarios: first, we adopt a phenomenological population model that does not evolve with redshift; second, we use the NF model to fit \bbh source-frame masses and merger redshift samples drawn from the redshift-evolving population. In the first scenario we measure a biased posterior for $H_0$ caused by the choice of the non-evolving population model that is not able to describe the redshift-evolving injected population. In the second case, we recover an unbiased measurement of $H_0$: the NF model was able to capture the redshift-evolution and correctly approximate the population model from the mock catalog.

\section{Application to a synthetic \bbh catalog: Calibration and analysis on GWTC-4.0} \label{bpop}

We now consider a synthetic astrophysical catalog generated with the \bpop pipeline \cite{bpop}, which provides an astrophysically motivated description of the \bbh population across cosmic time. Specifically, we use the fiducial \bpop catalog described in \cite{Bpopplace}. This catalog contains simulated \bbhs with primary mass reaching $\mathcal{O}(10^6) M_\odot$, which is far out of the LVK sensitivity \cite{gwtc4}. We consider only events within the detectable volume to correctly account for selection bias \cite{icarogw}. We cut out all the events at $z>2.5$: with this choice, we remove all the mergers with primary mass bigger than $\sim 200 M_\odot$. The population realization includes contributions from the isolated binary evolution formation channel and dynamically formed channels (globular clusters, nuclear clusters, and young clusters) \cite{mapellibbhformationchannels}. 
The \bpop population model predicts that \bbhs formed in the dynamical formation channel account for 34\% of the population while the ones from isolated formation channels account for 66\%.
For each channel, the source-frame primary and secondary masses, the merger redshift, together with the redshift merger rate $R(z)$ per unit comoving volume and source-frame time, are provided and correspond to one year. 
\bpop \cite{2019MNRAS.482.2991A,2020ApJ...894..133A,2023MNRAS.520.5259A,Bpopplace} generates mock \bbh populations in synthetic Universes that are uniquely defined by the cosmological star formation history of galactic fields and star clusters (young, globular, and nuclear clusters), the redshift evolution of stellar metallicity, the dynamical evolution of star clusters, the stellar evolution of single and binary stars, the natal mass and spin distribution, and spin alignment, of stellar BHs. In this work, we adopt the fiducial model described in Arca Sedda et al (2026). We refer the reader to the latest \bpop paper for more details about the code and the model.

\begin{figure}[ht!]
    \centering
    \includegraphics[width=0.5\columnwidth]{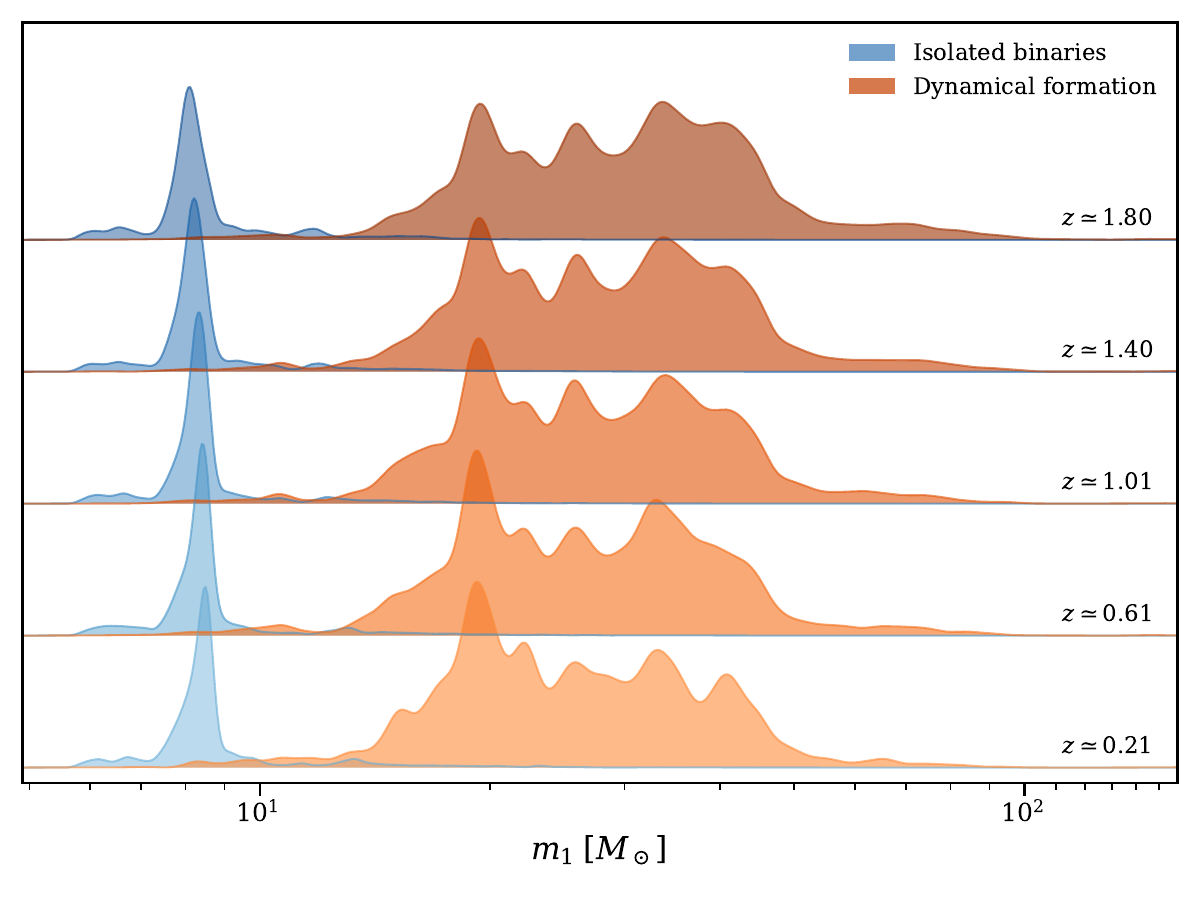}
    \caption{Redshift evolution of the primary \bbh mass distribution in the synthetic astrophysically simulated \bpop catalog, comparing the isolated binary channel (blue) and dynamical formation channel (orange), the latter represents 34\% of the population. Each curve shows the distribution of the primary source mass within a redshift bin. The population has a complex, multimodal structure and a strong redshift dependence. This complex behavior cannot be modeled with analytical phenomenological laws but can be captured using Normalizing Flows.}
    \label{bpopjoyplot}
\end{figure}

The \bpop population is characterized by a complex mass spectrum featuring sharp structures and a strong redshift evolution. Indeed, at different cosmic times, we expect different \bbh populations; thus, the redshift evolution is intrinsically embodied in the simulated catalog. This behavior can be seen in Fig.~\ref{bpopjoyplot}, which shows the evolution of the primary \bbh mass across redshift bins. Such complexity cannot be modeled using analytical phenomenological parametrizations, but it can be fitted using a Normalizing Flows model.
Before using a NF trained on \bpop to analyze GW events from GWTC-4.0, we perform a study with mock data to test the chosen architecture of the NF. 

\subsection{Calibration of the NF architecture} \label{calibration of the NF}

We generate a synthetic set of 818 GW observations from the \bpop catalog \cite{ferraiuologwsimulation, gwsimulation} in an O5-like era \cite{o5, ferraiuologwsimulation} by mapping the \bbh parameters to detector-frame quantities, fixing cosmology to $H_0 = 67.7 \; \text{km s}^{-1}\text{Mpc}^{-1}$ and $\Omega_m = 0.308$. We decided to use an O5-like scenario, as its increased sensitivity will allow us to detect a fraction of the \bpop population significantly larger than the one detectable in O4. As such, a NF calibrated for an O5-like observing scenario would suffice to approximate a population detectable in O4.

As a metric for the GW detection, we use a proxy for the signal-to-noise ratio (SNR) model defined as:
\begin{equation}
    \rho = \rho_0 \theta \left(\frac{\mathcal{M}_d}{\mathcal{M}_{d,0}}\right)^{5/6} \frac{d_{L,0}}{d_L} S \left(f_{\rm isco}(m_{1d}, m_{2d})\right),
    \label{snr}
\end{equation}
where $\mathcal{M}_d$ is the detector-frame chirp mass, $d_L$ is the luminosity distance, and $\theta$ encodes the geometric projection of the source relative to the detector network. The reference values $\mathcal{M}_{d,0}$ and $d_{L,0}$ are fixed according to \cite{ferraiuologwsimulation} to reproduce an O5-like detector network. 
The function $S \left(f_{\rm isco}(m_{1d}, m_{2d})\right)$ suppresses contributions from extremely massive \bbh mergers whose inspiral ends at frequencies below $\sim 20\,\mathrm{Hz}$. Such events can be found in the high-mass tail of the \bpop catalog and would otherwise produce artificially large SNRs; this effect is regularized by the function $S$, which downweights these events. 
The uncertainties on the measurement of the detector masses and luminosity distance for each event are generated using toy likelihood models as described in \cite{ferraiuologwsimulation}. We denote the set of 818 GW observations, and their error budgets on masses and distances, as parameter estimation samples. Using the same framework, we also generate a set of $10^8$ detectable GW signals to evaluate the selection effects as described in \cite{icarogw}.

\begin{figure}[ht!]
    \centering
    \includegraphics[width=0.6\columnwidth]{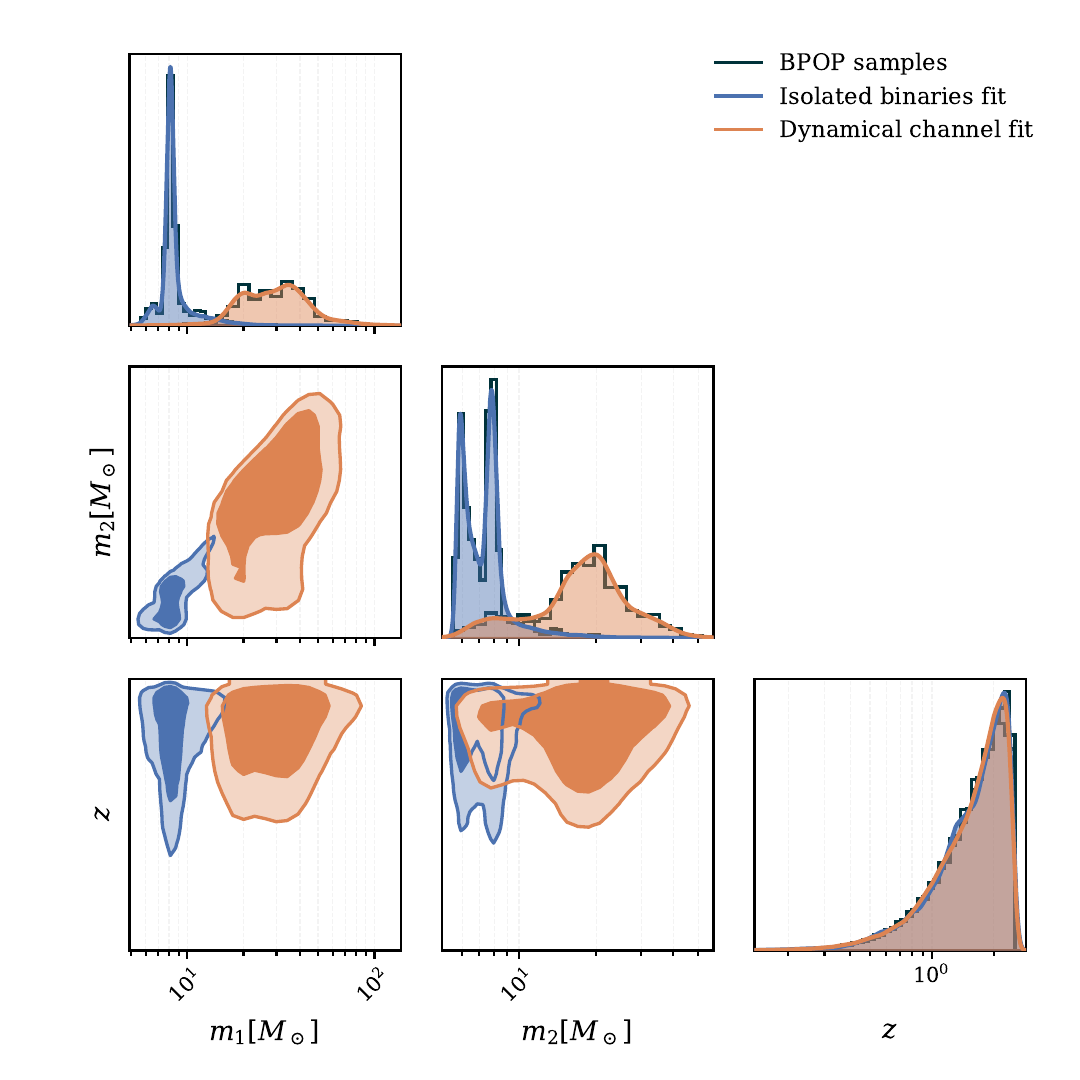}
    \caption{Corner plot showing the joint distribution of \bbh primary and secondary masses and merger redshift for the two formation channels. Blue and orange contours (that enclose the 68.3\% and 90\% credible regions) and smoothed kde curves show the Normalizing Flow fits of the isolated binary and dynamical channels, respectively. Grey histograms show the \bpop population samples and are shown only in the one-dimensional marginals. The Normalizing Flow fits correctly recover the mass and redshift spectra for both channels.}
    \label{bpopfit}
\end{figure}

For this analysis, we consider two effective formation channels on which we train two NFs: isolated binary evolution and dynamical formation in dense stellar environments. Each channel is modeled independently using a NF trained on the \bbh mass and redshift distributions. The details of the NF's architecture are given in App.~\ref{appendix nf architecture and training}.
Figure~\ref{bpopfit} shows the joint distributions of \bbh source-frame masses and merger redshift for the two formation channels and the corresponding normalizing flow fits of the \bpop population. The normalizing flow model correctly recovers the mass spectrum and its evolution with redshift.
We then define the detector-frame rate used for hierarchical Bayesian inference as:
\begin{equation}
     \frac{\mathrm{d}N}{\mathrm{d} d_L \mathrm{d}m_{1d} \mathrm{d}m_{2d} \mathrm{d}t_d} =
     \left[ (1-\phi)\, p_{\rm NF}^{\rm iso}(m_{1s}, m_{2s}, z)
     + \phi\, p_{\rm NF}^{\rm dyn}(m_{1s}, m_{2s}, z) \right]
     \frac{\mathrm{d}V_c}{\mathrm{d}z} \frac{1}{1+z} \frac{1}{|J_{d\rightarrow s}|}
     \frac{N_{\rm astro}}{T_{\rm obs}} .
     \label{eq detector rate bpop}
\end{equation}
Here, the observing time is set to $T_{\rm obs} = 1 \; \rm yr$ and the number of astrophysical \bbh mergers follows Eq~\ref{Nastro} and it is equal to $N_{\rm astro} = 6.6 \times 10^{4} \; \rm events/yr$, as predicted by \bpop. Further details on the detector rate are given in App.~\ref{appendix rate bpop}. In the above prescription, $\phi$ is the fraction of binaries formed in the dynamical formation channel.
Using Eq.~\ref{nexp}, we estimate the number of expected detections,  $N_{\rm exp} = 845 \pm 29$ , which is compatible within the 68.3\% credible interval with the number of 818 simulated parameter estimation samples. 

The simulated GW observations are used to perform hierarchical Bayesian inference on the Hubble constant $H_0$ and the dynamical formation channel mixture fraction $\phi$. We use a uniform prior for the Hubble constant $H_0 \sim \mathcal{U}(20,120)\;\mathrm{km\,s^{-1}\,Mpc^{-1}}$ and for the mixture fraction $\phi \sim \mathcal{U}(0,1)$, while the matter density parameter is fixed to \(\Omega_m = 0.308\) \cite{matterdensityparam}. The joint posterior distribution is shown in Fig.~\ref{h0bpop}(a): the injected values of $H_0$ and $\phi$ are recovered with no evidence for systematic biases. This demonstrates that the Normalizing Flows population model provides an accurate non-parametric representation of the fiducial \bpop catalog. Before testing this new population model on real GW data, we verified the reliability of the NF and ensured that the model was consistent and did not introduce biases in the $H_0$ estimation. This is further described in App.~\ref{appendix test wrong bpop}.

\subsection{Analysis on GWTC-4.0}

We apply the population model, validated on the simulated data, to real gravitational-wave observations from the GWTC-4.0 catalog, using the 153 detected events with inverse false alarm rate $\mathrm{IFAR} > 1\,\rm yr$ \cite{gwtc4}. To correct for selection biases, we use a set of software injections recovered by low-latency pipelines in real O3 and O4a data \cite{gwtc4inj}.
For the analysis, we use the rate model as written in Eq.~\ref{eq detector rate bpop}, where the two Normalizing Flows are the same as those benchmarked during the calibration. We firstly note that, using the reported $R(z)$ from \bpop, we would expect $N_{\rm exp} = 189 \pm 14$ detections, which are consistent with the size of the observed catalog within the 99\% credible interval. 

\begin{figure}
  \centering
  \begin{minipage}{0.48\linewidth}
    \centering
    \includegraphics[width=\linewidth]{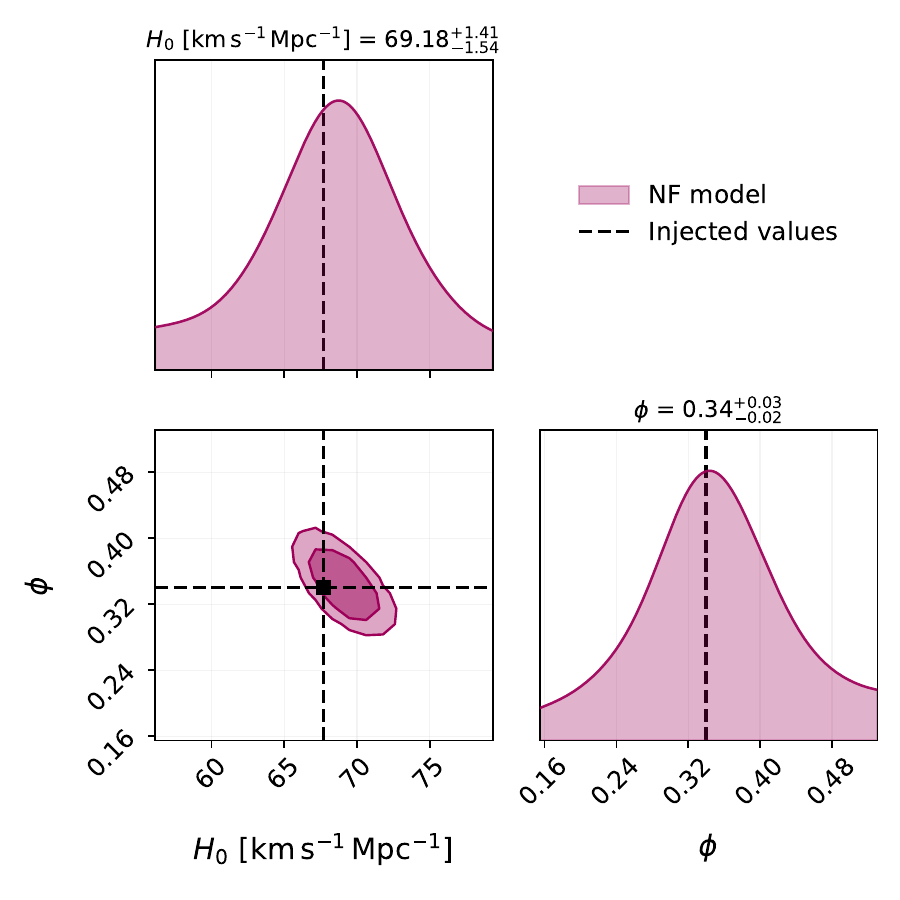}
    \textbf{(a)} Simulated GW observations
  \end{minipage}
  \hfill
  \begin{minipage}{0.48\linewidth}
    \centering
    \includegraphics[width=\linewidth]{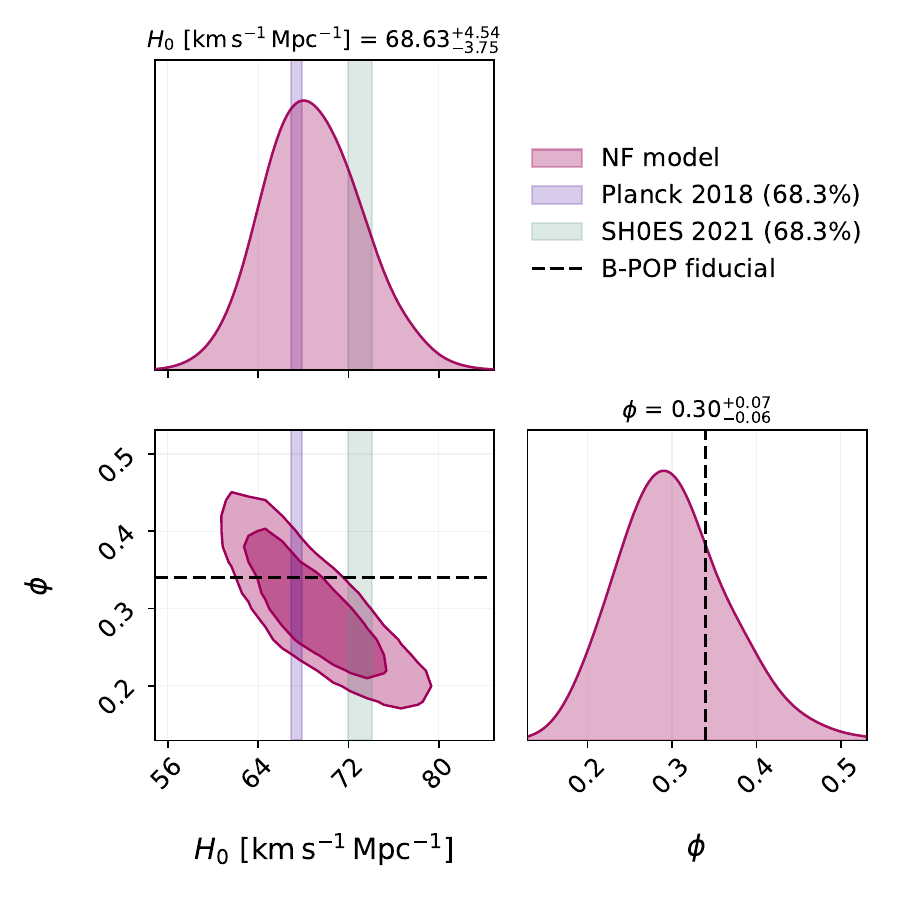}
    \textbf{(b)} GWTC-4.0 data
  \end{minipage}
  \caption{Joint posterior distributions of the Hubble constant $H_0$ and the mixture fraction of the dynamical \bbh formation channel $\phi$. (a) Inference performed on simulated GW data generated from the \bpop catalog using an MCMC sampler. Dashed lines indicate the injected values used to generate simulated data. Posteriors are compatible with the injected data, showing the correct fit of the Normalizing Flows to the \bpop population. (b) Constraints inferred from GWTC-4.0 data using a nested sampling algorithm within the same population model adopted for the simulated data. Vertical bands indicate the 68\% credible intervals from Planck 2018 (violet) and SHOES 2021 (green); the dashed line shows the \bpop fiducial value $\phi = 0.34$. In both panels, the contours enclose the 68.3\% and 90\% credible regions, and the diagonal panels show the marginalized one-dimensional posteriors.}
    \label{h0bpop}
\end{figure}

The joint posterior distribution inferred using GWTC-4.0 data is shown in Fig.~\ref{h0bpop}(b), it was obtained using a uniform prior for the Hubble constant $H_0 \sim \mathcal{U}(20,120)\;\mathrm{km\,s^{-1}\,Mpc^{-1}}$ and for the mixture fraction $\phi \sim \mathcal{U}(0,1)$, while the matter density parameter was fixed to \(\Omega_m = 0.308\) \cite{matterdensityparam}. The constraints on $H_0$ and $\phi$ are broader than those obtained from simulated observations because of the limited number of detected events and current detector sensitivity. The inferred value of $H_0=68.63^{+4.54}_{-3.75} \hu$ is consistent, within uncertainties, with both early- and late-Universe measurements \cite{plank, shoes}; the mixture fraction $\phi=0.30^{+0.07}_{-0.06}$ is compatible within the 68.3\% credible interval with the fiducial \bpop population, $\phi = 0.34$. 

From the posterior on $H_0$ and $\phi$, we also argue that \bpop provides a fair description of the mass and redshift distribution of currently detected GW events. This is a consequence of the fact that our posterior supports both the fiducial value of 0.34 of the mixture fraction between the two considered formation channels of \bbhs, as well as the current measurements of $H_0$ from early and late times. It is interesting to note that early-time values of $H_0$ are supported by a higher number of dynamically assembled \bbhs, while late-time values are supported by a lower number of them. This result stems from the fact that a population more dominated by dynamically formed \bbhs will display more binaries with high source mass, that can be obtained from detector masses only if the GW events are placed systematically at lower redshifts (lower $H_0$).

\begin{figure}[ht!]
    \centering
    \includegraphics[width=0.6 \columnwidth]{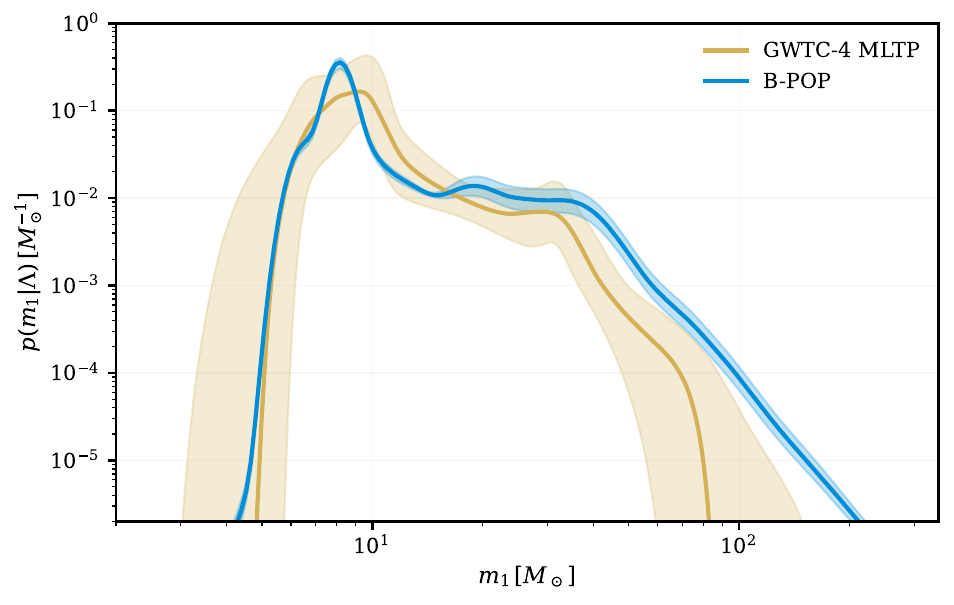}
    \caption{Primary source mass distribution $p(m_1)$, reconstructed from the fiducial GWTC-4.0 MLTP model \cite{gwtc4cosmo} and from \bpop. Solid lines indicate the median of the reconstructed distribution, while shaded regions show the 90\% C.I. obtained by varying the population parameters $\Lambda$ that, for the MLTP model correspond to the population parameters reported in \cite{gwtc4cosmo} while for \bpop they reduce to the mixture fraction $\phi$.}
    \label{m1_bpopvsgwtc4}
\end{figure}

\subsection{Caveats and discussion to interpret GWTC-4.0 results}

We note that the precision of $H_0$ that we obtain is significantly more precise than the one reported in the latest GWTC-4.0 cosmological studies \cite{gwtc4cosmo}. However, this is not surprising, as in our analysis, we are fixing the population model to the \bbh population described by \bpop, with the only additional degree of freedom being the mixture fraction between \bbhs formed in dynamical and isolated environments. In other words, we are ``fixing'' the features in the \bbh mass spectrum, which carries information on the source redshifts, to the ones predicted by \bpop. Instead, \cite{gwtc4cosmo} adopts general phenomenological models for the source mass that can account for different positions of the mass feature and therefore introduces more uncertainty in the determination of $H_0$. 
In other words, the NF model is less general than an analytical phenomenological model, but it can account for complex correlations in the mass-redshift space and provide guidance on how \bbh formed through isolated binary evolution and in dynamical environments contribute to the cosmological inference. Fig.~\ref{m1_bpopvsgwtc4} shows this behavior by comparing the reconstructed primary source-mass distribution from the GWTC-4.0 phenomenological Multi Peak (MLTP) model \cite{gwtc4cosmo} with the \bpop NF model: the fiducial MLTP model has narrower 90\% C.I. with respect to the \bpop model, because its only free parameter is $\phi$.

In this sense, $\phi$ and $H_0$ could also be understood as nuisance parameters for the inference that can give more flexibility to our rate model used for the inference. The effect of $\phi$ would be to adjust the ratio of local overdensities of \bbh in source mass, while the effect of $H_0$ would be to change how this mass and redshift spectrum is converted to detector masses and luminosity distance.
The most general and robust statement, from our results, is that \bpop reasonably fits the detected population of \bbhs if we assume $H_0$ closer to its early-time value, while it would need fewer \bbhs formed dynamically if $H_0$ is closer to its late-time value. 

We acknowledge that population models trained on NFs should be treated with care as they can introduce systematic biases on $H_0$ if they inherently mismatch observed data. To explore what would happen in these cases, in App.~\ref{appendix test wrong bpop} we perform a series of test cases using \bpop where we inferred $H_0$ and $\phi$ using NF models trained on mock catalogs that were not correctly describing observed data. We find that, when there are discrepancies in the most observed formation channels, the $H_0$ inference can be biased, but Bayes factors can be used to understand which NF fits the data better. As a consequence, to improve the robustness of the $H_0$ measure from this methodology, one should consider the results (and Bayes factors) obtained from several mock catalogs, as well as phenomenological models. We defer these extensive tests for future work.

Another important point to discuss relates to the circularity of the inference. \bpop obtains \bbh mergers redshift using the merger time of the binary and the redshift of formation (which is linked to the assumed Star Formation Rate). This procedure requires the assumption of a cosmological model to convert redshifts to lookback time and viceversa. For instance, a universe with very low values of $H_0$ would be very old and hence more binaries have time to merge, while the opposite is true for high values of $H_0$. We studied this effect by generating different \bpop catalogs simulated with different $H_0$ values and we analyzed the $H_0$ dependence of the detected \bbh primary masses and merger redshift. We noticed that the $H_0$ dependence is negligible in the merger redshift and that the detector mass spectrum at different $H_0$ values is compatible with the fiducial GWTC-4.0 model. The value of $H_0$ that we measure is driven by the source-frame mass features, which are rather independent of the fiducial cosmology assumed in the simulations. The details of this analysis are discussed in App.~\ref{BPOP with different H0}.

\section{Conclusions} \label{conclusions}

We introduced a non-parametric population framework based on Normalizing Flows to model the joint source-frame masses and merger redshift distributions of binary black holes in gravitational wave cosmology and population analyses. The model replaces standard phenomenological parametric prescriptions with non-parametric, data-driven density estimators that can capture non-trivial redshift-evolving populations and complex population features.

We validated the model using a redshift-evolving mock \bbh population and we showed that Normalizing Flows reconstruct the population. This benchmark demonstrates that the Normalizing Flows population model can be used to correctly perform hierarchical Bayesian inference.

We applied the model to a synthetic astrophysical \bbh catalog generated with the \bpop pipeline: the population is characterized by complex mass features and a strong redshift evolution. We modeled the isolated binaries and the dynamical formation channels with two independent and different Normalizing Flows, we combined them through a mixture coefficient and using synthetic GW observations we performed joint inference on the Hubble constant and on the dynamical formation channel fraction. We recovered the injected values without evidence for systematic biases, showing that Normalizing Flows can be used to describe astrophysically calibrated populations into the hierarchical Bayesian inference framework. 

Finally, we applied the population model to real gravitational wave data from the GWTC-4.0 catalog. We used the same Normalizing Flows model we used with simulated data without any modification. We obtained constraints on the Hubble constant and formation channel mixture fraction consistent with current cosmological measurements and astrophysical expectations.

This work demonstrates that non-parametric, astrophysically calibrated population models based on Normalizing Flows provide a consistent model, with the correct caveats, for gravitational wave population inference. 

\section{Acknowledgments}
We are grateful to M.~Spera for discussions on the project. G.~Antinozzi gratefully acknowledges SISSA for the Ph.D. scholarship and thanks M.~Spera and A.~Lapi for their insightful feedback as supervisors.
S.M and A.S. are supported by ERC grant GravitySirens  101163912. Funded by the European Union. Views and opinions expressed are however those of the author(s) only and do not necessarily reflect those of the European Union or the European Research Council Executive Agency. Neither the European Union nor the granting authority can be held responsible for them. This material is based upon work supported by NSF's LIGO Laboratory which is a major facility fully funded by the National Science Foundation. F.S. has been funded by the European Union –NextGenerationEU under the Italian Ministry of University and Research (MUR) "Decreto per l’assunzione di ricercatori internazionali post-dottorato PNRR" - Missione 4 "Istruzione e Ricerca" Componente 2 "Dalla Ricerca all’Impresa" del PNRR - Investimento 1.2 “Finanziamento di progetti presentati da giovani ricercatori” - CUP D13C25000700001. M.A.S. acknowledges funding from the European Union’s Horizon 2020 research and innovation programme under the Marie Skłodowska-Curie grant agreement No.~101025436 (project GRACE-BH) and from the MERAC Foundation through the 2023 MERAC prize. M.A.S. acknowledges the ACME project which has received funding from the European Union's Horizon Europe Research and Innovation programme under Grant Agreement No.~101131928.

\section{Data availability statement}
The data that support this work are available at: \url{https://doi.org/10.5281/zenodo.22658639}.

\appendix

\section{Normalizing Flows architecture and training} \label{appendix nf architecture and training}
In this appendix we provide the details of the implementation of the Normalizing Flows model introduced in Sec.~\ref{HBI e NF}. We describe the data preprocessing, model architecture, training procedure and model selection criteria.

In order to enhance the training procedure of the \textsc{zuko} \cite{zuko} Normalizing Flows model, the input population samples are normalized to a fixed bounded domain $[-b,b]$. Given the set of input samples $ x = (m_{1s}, m_{2s}, z)$, we compute the minimum $x_{\rm min}$ and maximum $x_{\rm max}$ values of each dimension from the training dataset (80\% of the total dataset). Each component is rescaled as follows:
\begin{equation}
     x_{\rm norm} = \frac{2b}{x_{\rm max} - x_{\rm min}} (x - x_{\rm min}) - b .
\end{equation}
This choice is made to map the normalized input data within the domain over which the rational-quadratic splines of the \textsc{zuko} library are defined \cite{nsf}. Outside this domain, the Normalizing Flow does not correctly model the input distribution. In this work, we fix $b = 3.5$. The corresponding Jacobian correction is consistently taken into account.

We use \textsc{zuko} Neural Spline Flows with a standard 3-dimensional Gaussian latent distribution. The main parameters of the flow are the number of transformation layers and the number of hidden neurons in the neural network.  

We train the model by minimizing the negative log-likelihood (NLL) \cite{nfinference} of the training data using the Adam \cite{adam} optimizer, with a fixed learning rate of 0.003 and a fixed batch size. An early stopping method is used on the validation set (20\% of the total dataset) to prevent overfitting. 

\begin{figure}[htpb]
\centering

\begin{minipage}{\linewidth}
    \centering
    \includegraphics[width=0.8\linewidth]{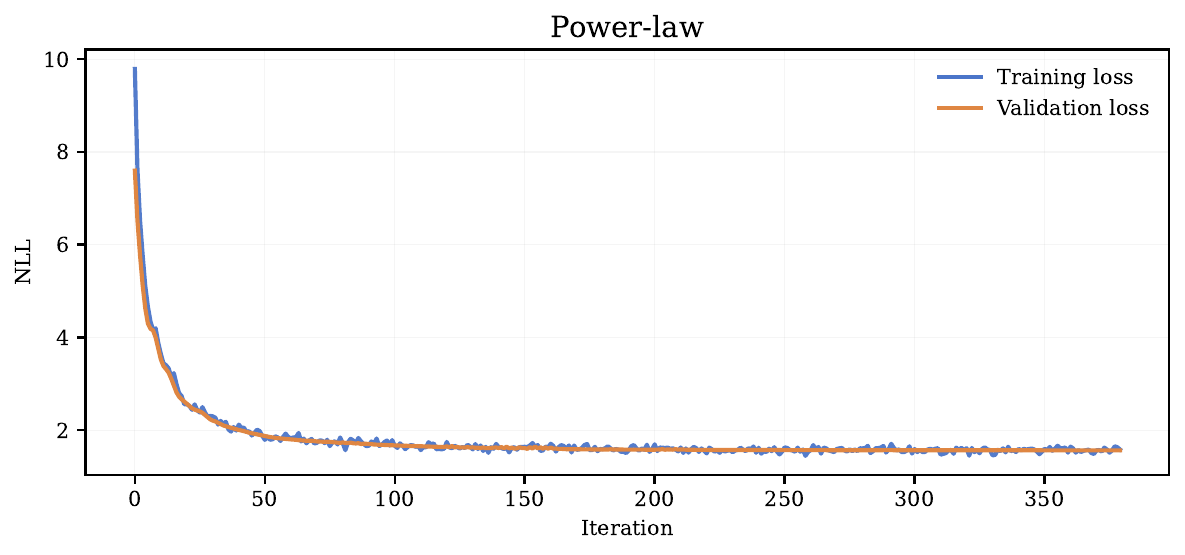}
\end{minipage}
\begin{minipage}{\linewidth}
    \centering
    \includegraphics[width=0.8\linewidth]{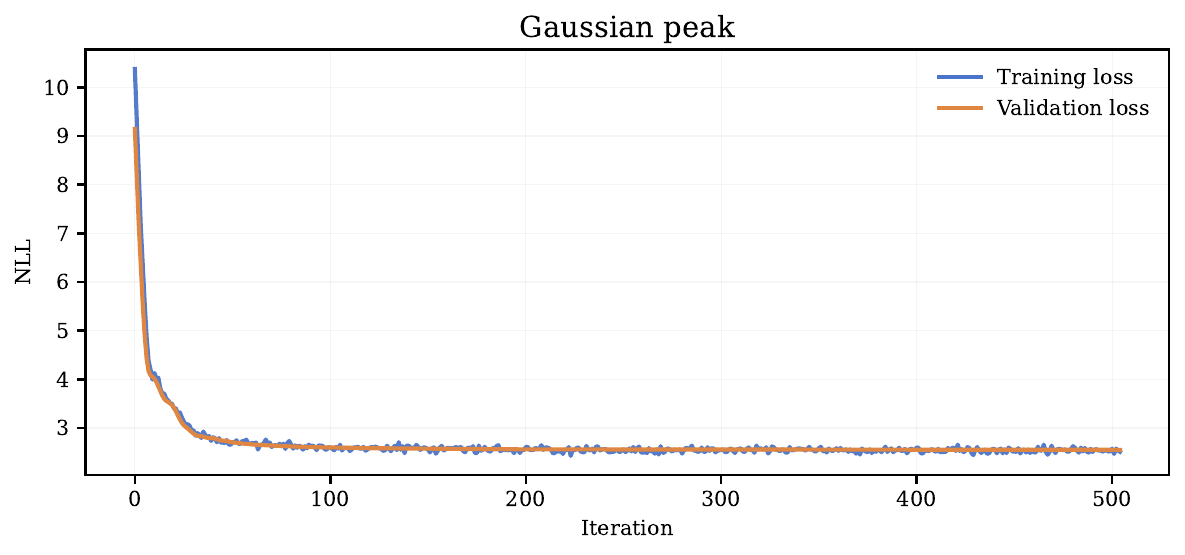}
\end{minipage}
\begin{minipage}{\linewidth}
    \centering
    \includegraphics[width=0.8\linewidth]{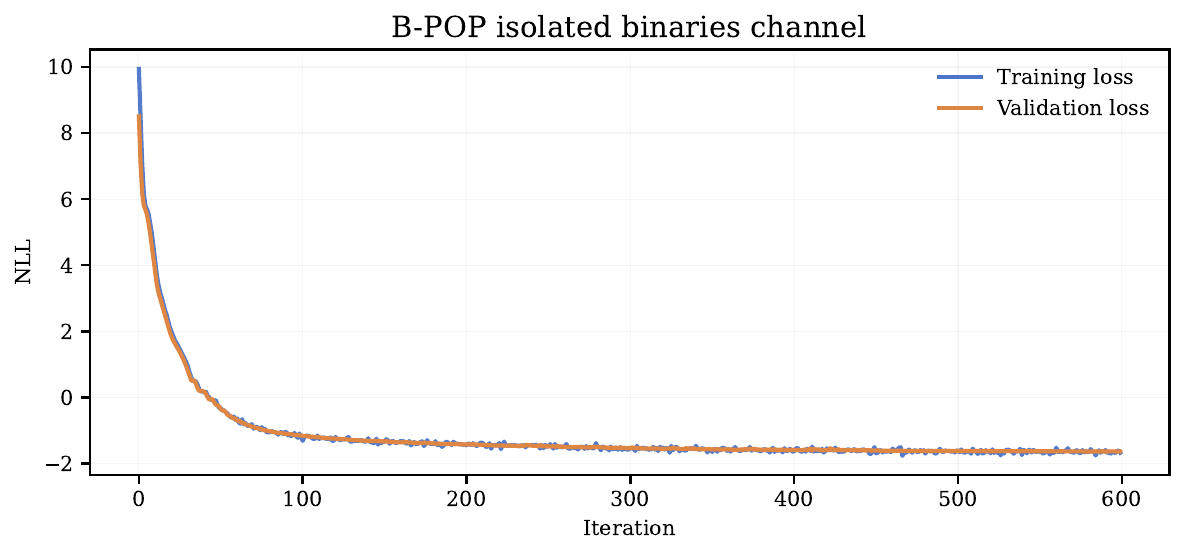}
\end{minipage}
\begin{minipage}{\linewidth}
    \centering
    \includegraphics[width=0.8\linewidth]{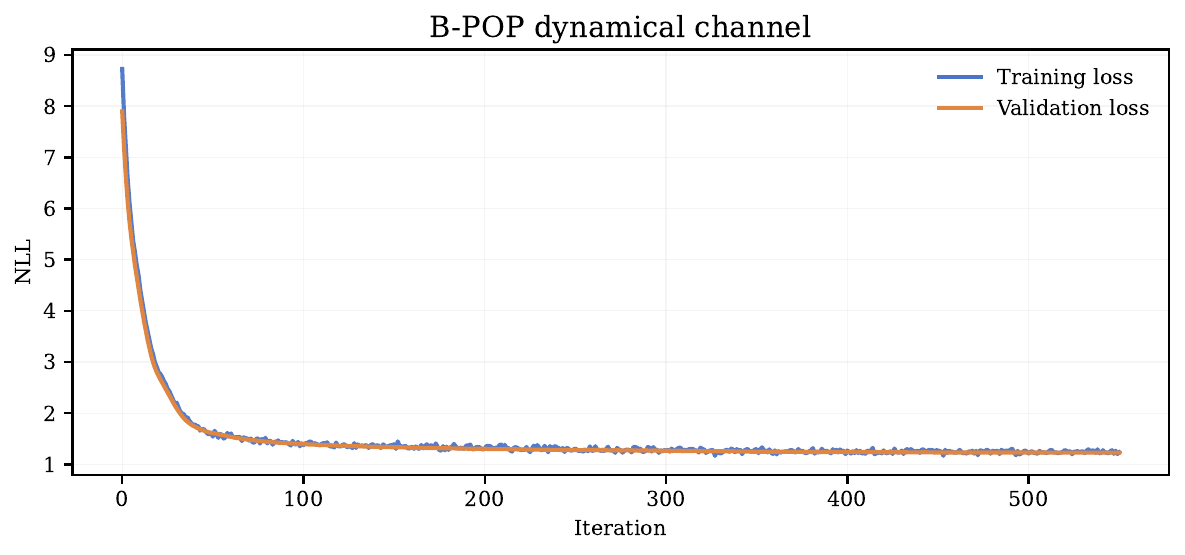}
\end{minipage}
\caption{Training and validation losses as a function of the training iterations for the four population models considered in this work: the mock redshift-evolving Power Law and Gaussian Peak, the \bpop isolated binaries channel and the \bpop dynamical channel.}
\label{trainingcurves}
\end{figure}

A summary of the Normalizing Flows architectures we used for all the different population catalogs are reported in Tab.~\ref{nf arch} and the training and validation loss curves are shown in Fig.~\ref{trainingcurves}, where we observe a stable convergence behaviour and no overfitting.

\begin{table}[h!]
\centering
\begin{tabular}{l|c|c|c|c|c}
\hline
 & Layers & Hidden neurons  & Batch size & Train loss & Validation loss \\
\hline
Power Law & 4 & 10  & 1000 & 1.5665 & 1.5663 \\
Gaussian Peak & 3 & 15  & 1000 & 2.5470 & 2.5513 \\
isolated binaries channel & 3 &  10 & 2048 &  -1.6492 &  -1.6387 \\
dynamical channel &  3 &  10 & 2048  &  1.2426 & 1.2299 \\
\hline
\end{tabular}
\caption{Normalizing Flow architectures and training parameters used for all the population models we used in this work, in all models, the learning rate and the number of iterations are respectively fixed to 0.003 and 600.}
\label{nf arch}
\end{table}

\newpage
\section{Benchmark of the model on a mock catalog} \label{appendix mock catalog benchmark}

We simulate a mock catalog of \bbhs from the \textsc{Power Law + Evolving Peak} model as the one employed in \cite{pierrasystematics}.  The primary mass distribution follows a Power Law and a truncated Gaussian Peak, whose mean $\mu$ evolves with redshift
\begin{equation}
    \mu(z) = \mu_{z_0} + \mu_{z_1} z .
\end{equation}
where $\mu_{z_0}$ and $\mu_{z_1}$ respectively represent the mean of the Gaussian Peak at $z=0$ and at $z=1$. The low-mass end of the distribution is smoothed over a scale \(\delta_m\). The secondary mass is drawn conditionally on the primary mass from a smoothed Power Law distribution. The full joint probability density can be written as
\begin{equation}
p(m_1, m_2, z|\Lambda) = p(m_1 | z, \Lambda)\, p(m_2 | m_1, \Lambda)\, p(z|\Lambda).
\end{equation}
The merger redshift distribution follows the Madau-like rate evolution \cite{madaurate, icarogw}.
The population parameters used to generate the mock catalog are chosen to be in agreement with those inferred from the GWTC-3.0 analyses \cite{gwtc3pop, gwtc3cosmo} and are reported in Tab.~\ref{mock param}.

\begin{table}[h!]
\centering
\begin{tabular}{l l l}
\hline
Parameter $(\Lambda)$ & Definition & Value \\
\hline
\multicolumn{3}{l}{\textbf{Cosmology (flat $\Lambda$CDM; fixed in the mock)}} \\
$H_0$ &Hubble constant [$\mathrm{km\,s^{-1}\,Mpc^{-1}}$] & $67.7$ \\
$\Omega_m$ & Matter density parameter &$0.308$ \\
\hline
\multicolumn{3}{l}{\textbf{Merger redshift distribution $p(z)$ (Madau-like rate evolution)}} \\
$\gamma$ & Low-$z$ evolution slope &$4.59$ \\
$\kappa$ & High-$z$ evolution slope & $2.86$ \\
$z_p$ & Redshift of Peak rate & $2.47$ \\
\hline
\multicolumn{3}{l}{\textbf{Primary mass distribution $p(m_{1s}\mid z)$ (Power Law + Evolving Peak})} \\
$\alpha$ & Power Law index of the primary mass & $4.09$ \\
$m_{\rm min}$ & Minimum source-frame mass [$M_\odot$] & $4.98$ \\
$m_{\rm max}$ & Maximum source-frame mass [$M_\odot$] & $112.5$ \\
$\mu_{z_0}$ & Gaussian Peak mean at $z=0$ [$M_\odot$] & $32.27$ \\
$\mu_{z_1}$ & Linear redshift coefficient of the peak mean [$M_\odot$] &
$2.5$ \\
$\sigma_{z_0}$ &Gaussian Peak width [$M_\odot$] &$3.88$ \\
$\phi$ & Fraction of events in the Gaussian Peak &
$0.08$ \\
$\delta_m$ & Low-mass smoothing scale [$M_\odot$] &$3.70$ \\
\hline
\multicolumn{3}{l}{\textbf{Secondary mass distribution $p(m_{2s}\mid m_{1s})$}} \\
$\beta$ & Power Law index of the secondary mass &$-1.23$ \\
\hline
\end{tabular}
\caption{Population parameters used to generate the redshift-evolving mock \bbh population.}
\label{mock param}
\end{table}

\begin{figure}[ht!]
    \centering
    \includegraphics[width=0.4\columnwidth]{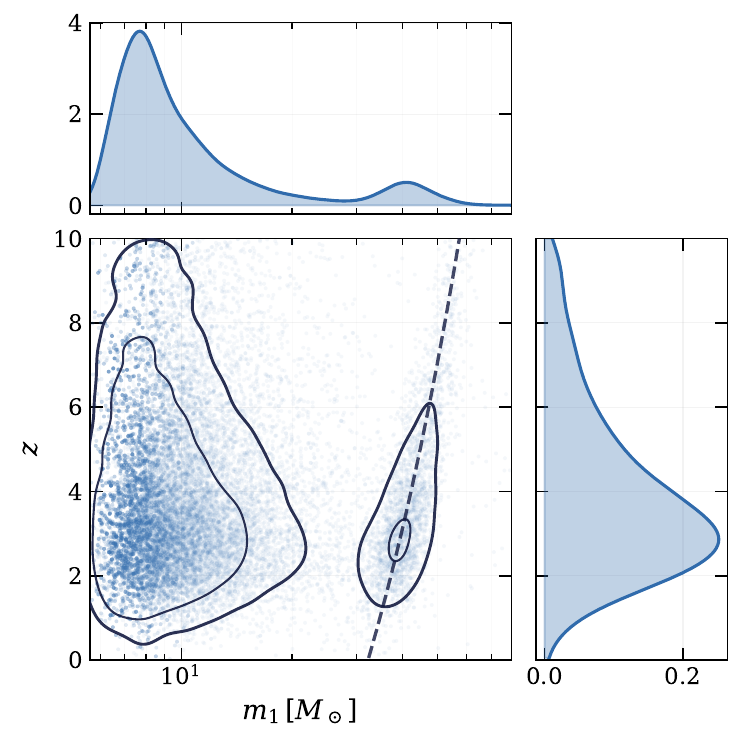}
    \caption{Astrophysical distribution of the mock \bbh source-frame masses and merger redshifts drawn from the \textsc{Power Law + Evolving Peak} population, used to benchmark the model. The plot is shown in the $\log m_1 - z$ plane, where the dots are samples drawn from the redshift-evolving distribution. Contour lines enclose the 68\% and 90\% density regions. The dashed line shows the linear redshift evolution of the mean of the Gaussian Peak: $\mu(z) = z \; \mu_{z_1} + \mu_{z_0}$. The top and right panels show the corresponding marginal distributions.}
    \label{mockpop}
\end{figure}

The astrophysical distribution of primary masses and merger redshift in the mock catalog is shown in Fig.~\ref{mockpop}, where the linear evolution of the Gaussian Peak is clearly visible. The fraction of GW events in the Evolving Peak is set to $\phi=0.08$, following Eq.~\ref{eq:mix}. We note that, throughout this benchmark, the Power Law and Evolving Peak components should be interpreted as phenomenological representatives of two distinct formation channels and are combined through the mixture coefficient $\phi$. However, in standard GW analyses \cite{gwtc4cosmo,gwtc4pop}, $\phi$ has no direct physical interpretation in terms of formation channels and it simply parametrizes the fraction of events in the peak.

Mock GW observations are generated by reweighting a set of detected injections in O3 detector data \cite{gwtc3} according to a \textsc{Power Law + Evolving Peak} population. The cosmological parameters are fixed to $H_0 = 67.7 \; \text{km s}^{-1}\text{Mpc}^{-1}$ and $\Omega_m = 0.308$ \cite{matterdensityparam}. This procedure consistently accounts for GW selection effects under the assumed population model. 
For this benchmark, we assume that we can perfectly measure the detector masses and luminosity distance of the GW source. This is an unrealistic assumption for real data, but it is a useful benchmark to control, even in the most optimistic scenario, if NF can be employed for redshift-evolving mass distributions.

\begin{figure}[ht!]
    \centering
    \includegraphics[width=0.5\columnwidth]{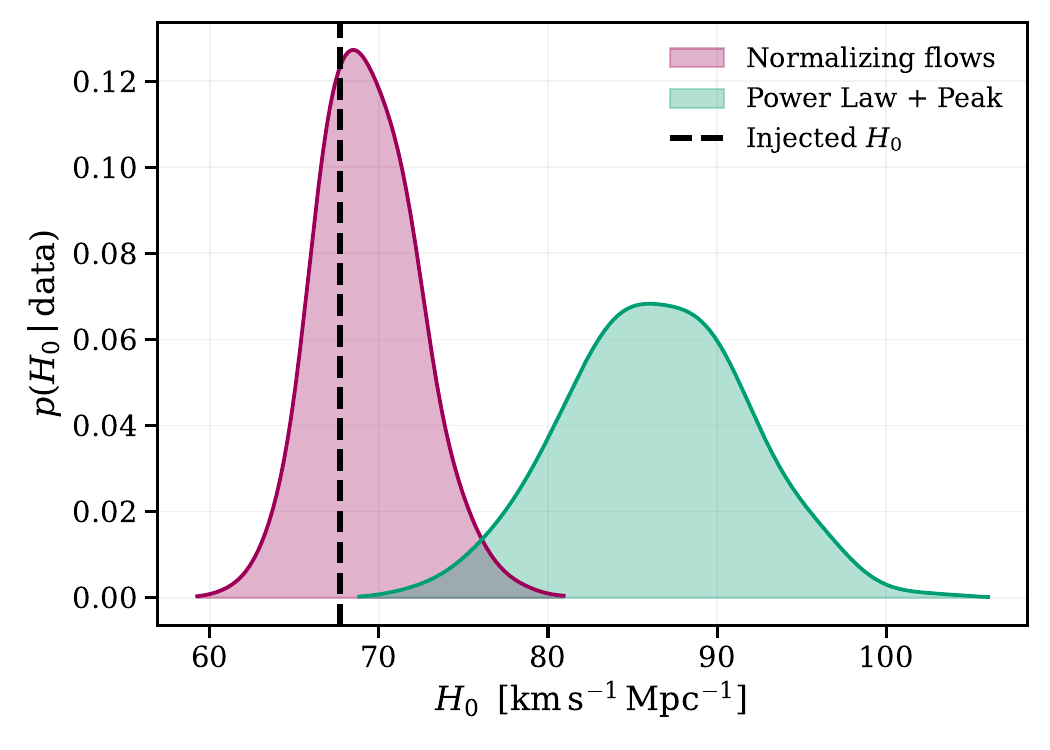}
    \caption{Posterior distributions of $H_0$ inferred using the mock GW catalog under two different population assumptions. The purple posterior is obtained when masses are modeled with the phenomenological \textsc{Power Law + Peak} that does not evolve with redshift: the injected value of $H_0$ (vertical dashed line) falls outside the 99.7\% C.I. of the inferred posterior. The magenta posterior shows inference results obtained using Normalizing Flows to fit the redshift-evolving population: the evolution is well captured and the injected $H_0$ value is recovered.}
    \label{H0mockposteriors}
\end{figure}

In the first test case, we model the mass distribution with a \textsc{Power Law + Peak}, thus ignoring the redshift evolution. We infer $H_0$ using the Markov chain Monte Carlo (MCMC) technique, fixing the matter density parameter to $\Omega_m = 0.308$ and using a uniform prior on the Hubble constant $H_0 \sim \mathcal{U}(20,120)\;\mathrm{km\,s^{-1}\,Mpc^{-1}}$. We fix the rest of the population parameters to their true values reported in App.~\ref{appendix mock catalog benchmark}, setting the redshift-evolution parameter to $\mu_{z_1}=0$.
The $H_0$ posterior distribution is shown in Fig.~\ref{H0mockposteriors} and reproduces the systematic bias already reported in the literature \cite{pierrasystematics},  $H_0$ can be found at the 99.7\% interval. The analytical form of the \textsc{Power Law + Peak} phenomenological model, which does not evolve with redshift, is the origin of the bias. This implies that whenever a redshift-evolving population is modeled using a population that is fixed in redshift, a systematic error arises from the mismatch between the true population and the phenomenological approximation.
In contrast, as we can see from Fig.~\ref{H0mockposteriors}, NFs are capable of approximating the catalog of GW sources and recover a posterior that includes the simulated $H_0$ value.

\begin{figure}[ht!]
    \centering
    \includegraphics[width=0.6\columnwidth]{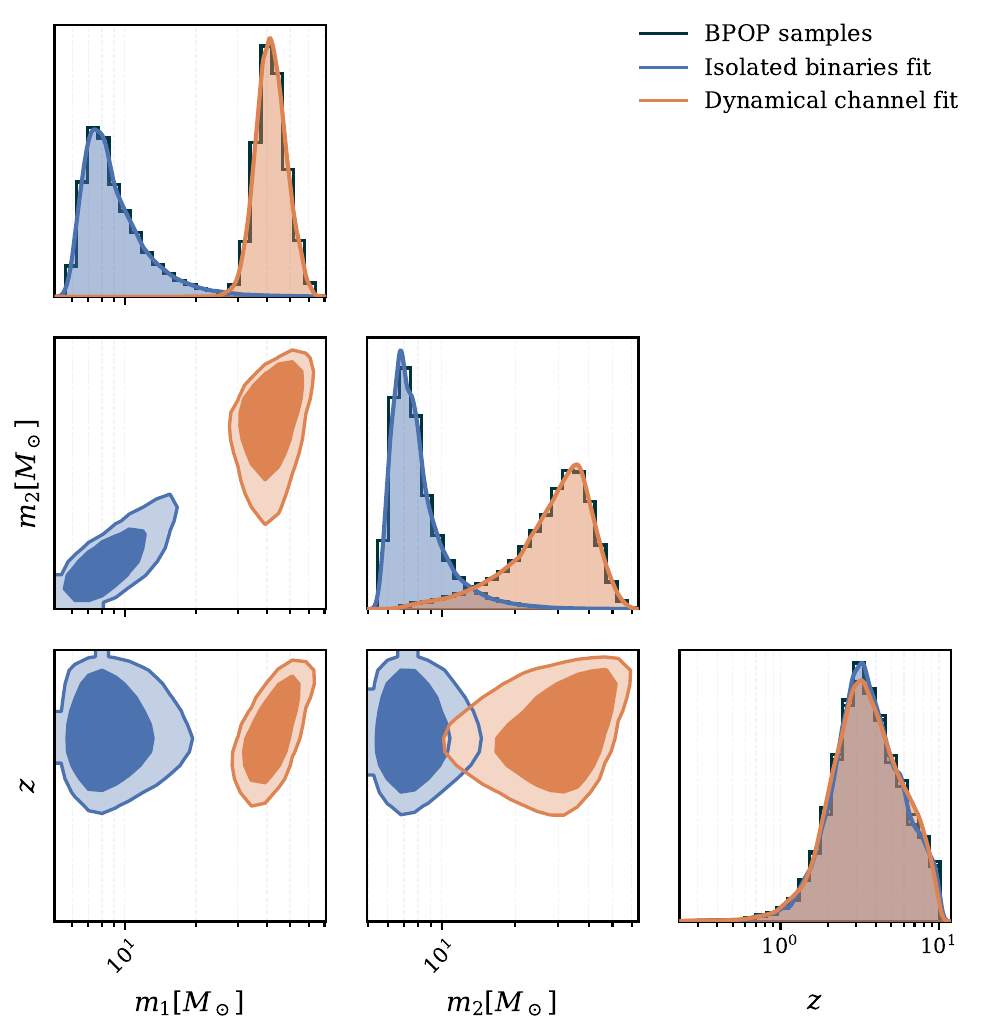}
    \caption{Corner plot showing the joint distribution of \bbh masses and merger redshift for the two mock populations. Blue and orange contours (enclosing the 68.3\% and 90\% credible regions) and smoothed KDE curves represent the normalizing flow fits for the Power Law and the Evolving Peak, respectively. Grey histograms, shown only in the one-dimensional marginals, correspond to samples drawn from the mock redshift-evolving catalog. The Normalizing Flows accurately recover the underlying mass and redshift distributions for both channels.}
    \label{plpeakfit}
\end{figure}

In the second test case, we model the mass using two different NFs linked through a mixture fraction $\phi$ as in Eq.~\ref{eq:mix}. The first population is composed of all the GW events generated from the Power Law, while the second is composed of the GW events generated by the Redshift-Evolving Peak, both populations and their NF fits are shown in Fig.~\ref{plpeakfit}.
As in the previous case, we infer $H_0$ fixing $\Omega_m = 0.308$ and the mixture coefficient to its injected value $\phi = 0.08$ and assuming a uniform prior on the Hubble constant $H_0 \sim \mathcal{U}(20,120)\;\mathrm{km\,s^{-1}\,Mpc^{-1}}$. Results are shown in Fig.~\ref{H0mockposteriors}, the systematic bias on $H_0$ is fully corrected thanks to the Normalizing Flows, which accurately fit the redshift-evolving population.

\newpage

\section{Detector-frame rate for \bpop catalogs} \label{appendix rate bpop}

In this appendix we discuss Eq.~\ref{eq detector rate bpop} which describes the detector-frame merger rate that uses the astrophysical synthesis catalogs generated with the \bpop pipeline and fitted with the Normalizing Flows model.

The \bpop catalogs provide \bbh redshift $z$ samples drawn from a distribution proportional to:
\begin{equation}
R(z)\,\frac{dV_c}{dz}\,\frac{1}{1+z},
\label{merger redshift samples eq}
\end{equation}
where $R(z)$ is the merger rate density per comoving volume and source-frame time. This implies that, when Normalizing Flows are trained directly on the catalog samples, the resulting $p_{\rm NF}(m_{1s}, m_{2s}, z)$  is normalized with respect to this sampling measure. Thus, an explicit normalization is needed in order to consistently compute Eq.~\ref{cbcrate}. This leads to the parametrization of the detector-frame rate for \bpop catalogs in Eq.~\ref{eq detector rate bpop}, in which the total number of astrophysical mergers happening in the Universe during the observation time $T_{\rm obs}$ is defined as:

\begin{equation}
N_{\rm astro} =T_{\rm obs}\int dz \;R(z)\,\frac{1}{1+z}\,\frac{dV_c}{dz}.
\label{Nastro}
\end{equation}
\newpage

\section{Test of consistency of the Normalizing Flow model}\label{appendix test wrong bpop}
In this appendix we perform a test to check the consistency and reliability of the NF model used in the HBI framework and to understand if it raises systematic biases in the inference process.
We use mock simulated data from the \bpop catalog, as explained in Sec.~\ref{calibration of the NF}. We perform inference of $H_0$ and $\phi$ with three different NF training assumptions and for each of them we measure the Bayesian evidence $\ln Z$:
\begin{enumerate}
    \item NF trained on the full \bpop population (isolated binaries and dynamical binaries from globular, nuclear and young clusters). We measure $\ln Z_1 = 1968.04 \pm 0.29$ and both $H_0$ and $\phi$ are recovered within their 68.3\% C.I.
    \item NF trained on the dynamical channel only, neglecting the isolated channel. We obtain $\ln Z_2 = 1941.97 \pm 0.37$, $H_0$ is recovered within its 68.3\% C.I.
    \item NF trained on the isolated channel and only the nuclear and young clusters of the dynamical population, neglecting the globular clusters. We measure $\ln Z_3 = 1931.31 \pm 0.44$ and both $H_0$ and $\phi$ are biased by up to $10 \sigma$.
\end{enumerate}
The correct model (case 1) has the highest evidence and the resulting Bayes factors strongly disfavor both mismodeled cases. For the case 2, $H_0$ remains unbiased despite the missing isolated channel, but the evidence clearly shows that this model is not favoured. Thus, the evidence is sensitive to the population mismatch even when it does not translate into a visible bias on $H_0$. In case 3 the posteriors on $H_0$ and $\phi$ are strongly biased because the population that we removed are the globular clusters, which dominate the dynamical channel and populate the primary masses around $\sim 35 M_\odot$, where the majority of GW events lie. Removing globular clusters  significantly distorts the mass spectrum in the observationally relevant range for the measurement of $H_0$, causing a strong bias. Instead, removing the isolated channel leaves this range mostly unaffected, so $H_0$ is not biased.

This test shows that NFs used in the HBI framework are sensitive to population change. It also shows that recovering an unbiased $H_0$ is not sufficient evidence that the assumed population model is correct, it should be always checked also with the Bayesian evidence.

\section{\bpop catalog simulated with different $H_0$} \label{BPOP with different H0}
\begin{figure}[htpb]
  \centering
  \begin{minipage}{0.48\linewidth}
    \centering
    \includegraphics[width=\linewidth]{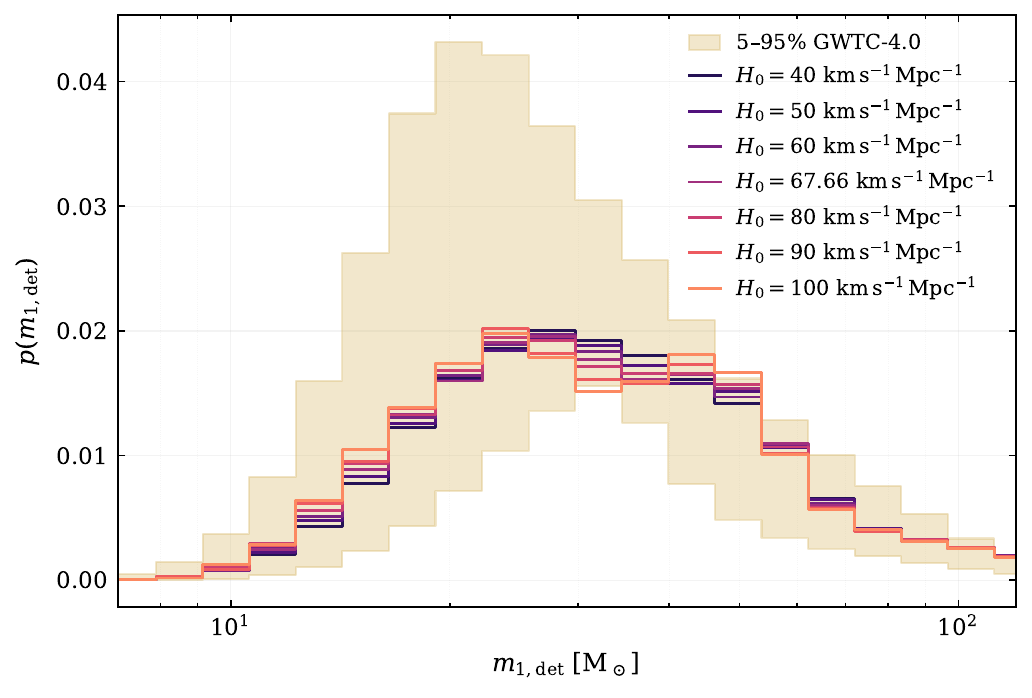}
    \textbf{(a)}
  \end{minipage}
  \hfill
  \begin{minipage}{0.48\linewidth}
    \centering
    \includegraphics[width=\linewidth]{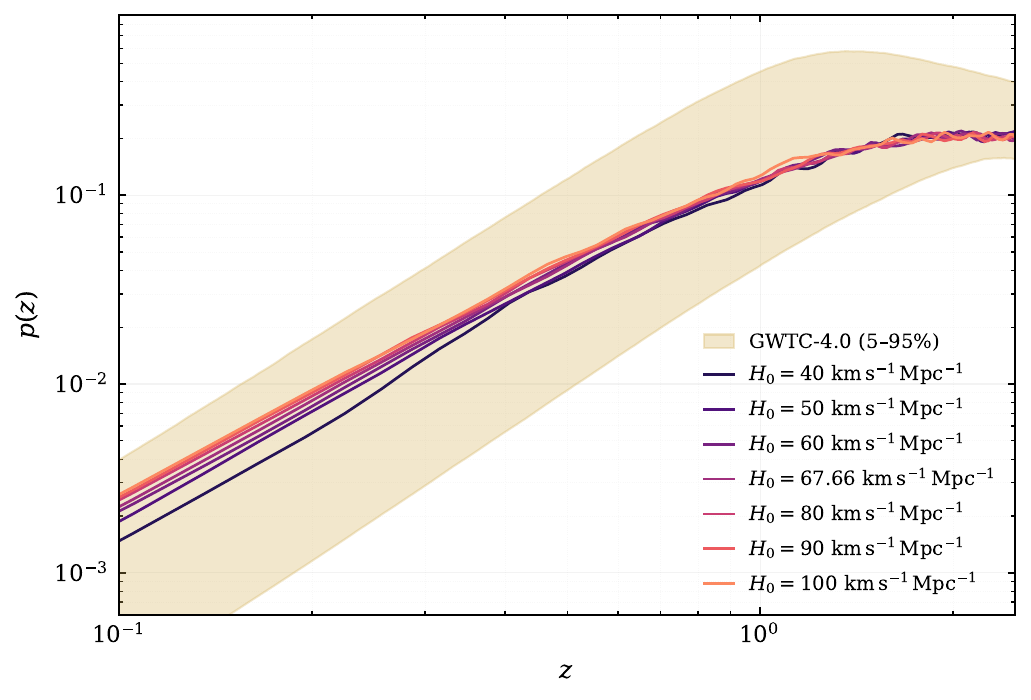}
    \textbf{(b)}
  \end{minipage}

  \vspace{0.3cm} 

  \begin{minipage}{0.48\linewidth}
    \centering
    \includegraphics[width=\linewidth]{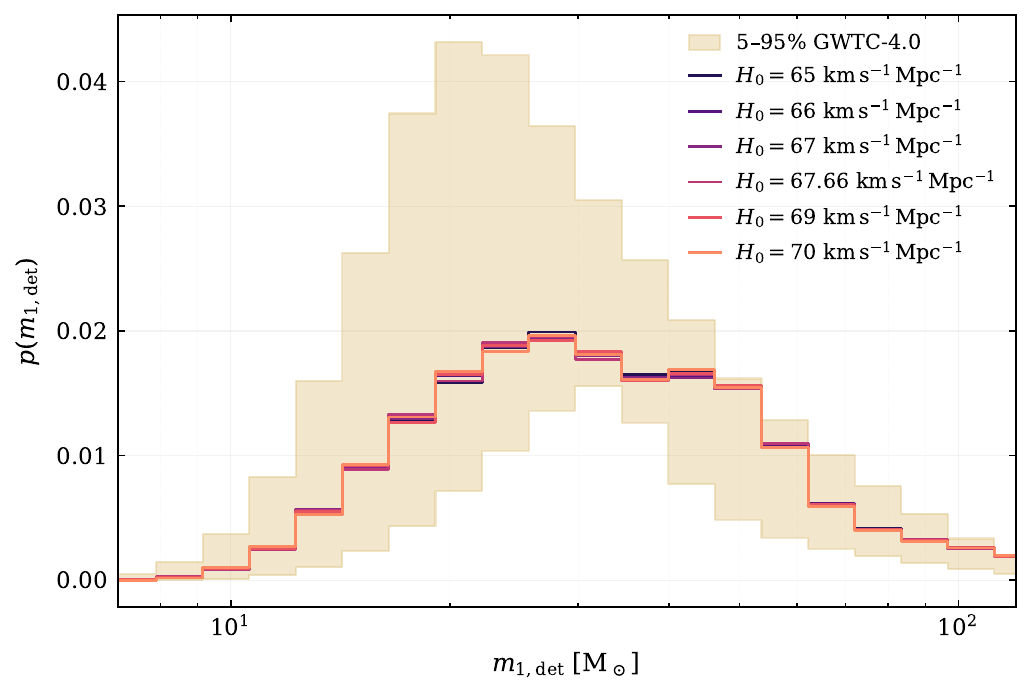}
    \textbf{(c)}
  \end{minipage}
  \hfill
  \begin{minipage}{0.48\linewidth}
    \centering
    \includegraphics[width=\linewidth]{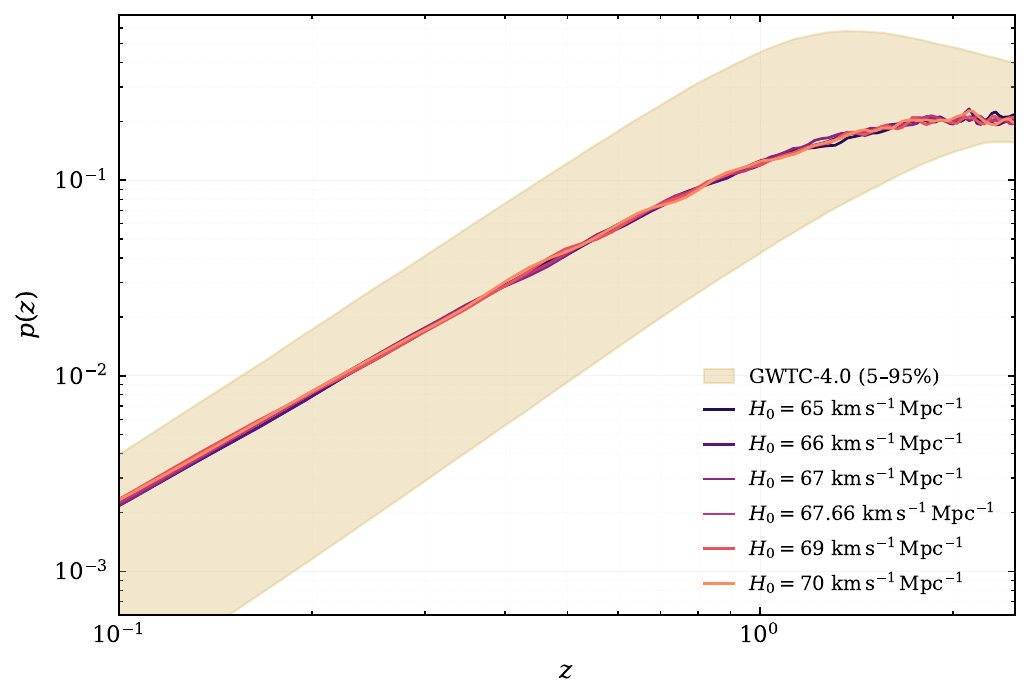}
    \textbf{(d)}
  \end{minipage}

  \caption{\bpop catalogs generated assuming large $H_0$ variations $H_0 \in \{40,50,60,67.7,80,90,100\}\;\hu$ (top) and small $H_0$ variations (bottom). Panels (a) and (c) show detector-frame primary mass histograms. Panels (b) and (d) show the normalized merger redshift $R(z)$ distribution. In all panels solid lines represent the different distributions when varying $H_0$, and the shaded regions represent the 90\% C.I. of the fiducial GWTC-4.0 MLTP model.}
  \label{fig H0 variation plots}
\end{figure}

\begin{figure}[htpb]
  \centering
  \begin{minipage}{0.48\linewidth}
    \centering
    \includegraphics[width=\linewidth]{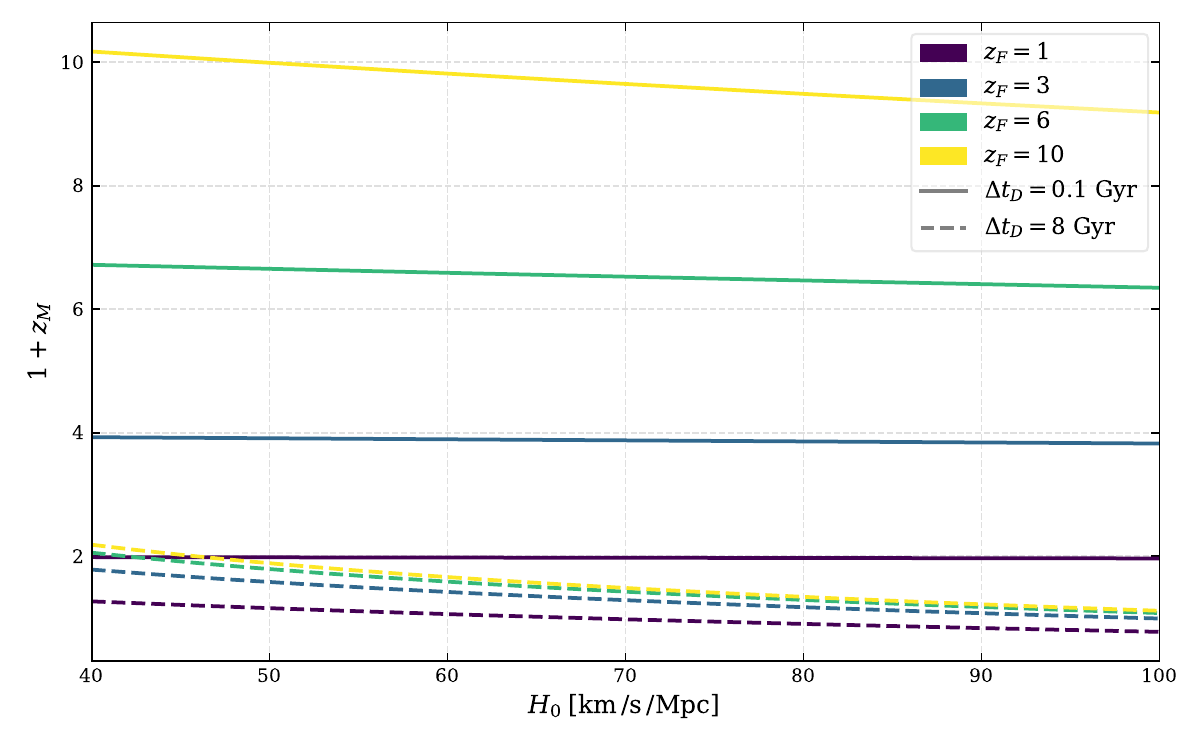}
    \textbf{(a)}
  \end{minipage}
  \hfill
  \begin{minipage}{0.48\linewidth}
    \centering
    \includegraphics[width=\linewidth]{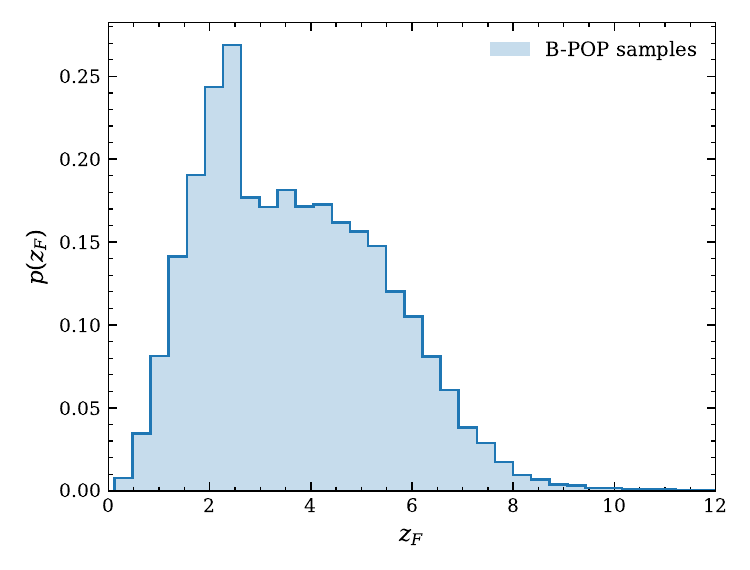}
    \textbf{(b)}
  \end{minipage}

  \caption{\textbf{(a)} $z_M$ as a function of $H_0$ for different values of $z_F$ and $\Delta t_D $. The dependence on $H_0$ is negligible for $z_F<6$ and short $\Delta t_D$. It becomes more visible for $z>6$ and longer $\Delta t_D$. \textbf{(b)} $z_F$ samples from the \bpop catalog we considered: $\sim 90\%$ of the \bbh mergers have $z_F<6$.}
  \label{H0vszm}
\end{figure}

In this appendix, we analyze whether the assumption of a fixed cosmology when simulating \bpop catalogs introduces circularity in our analysis. In particular, we want to understand how sensitive the astrophysical catalog is to the assumed value of $H_0$. 
In the \bpop pipeline, $H_0$ is needed to estimate the merger redshift $z_M$ of the sources:
\begin{equation}
    z_M = t^{-1}_{LB}[t_{LB}(z_F)-\Delta t_D],
    \label{merger redshift eq}
\end{equation}
where $t_{LB}$ is the lookback time that depends on $H_0$, $z_F$ is the formation redshift and $\Delta t_D$ is the delay time.
We generate several \bpop realizations varying only the cosmology used to convert the lookback time into merger redshift, keeping all the other astrophysical prescriptions fixed. We choose a set of values close to the fiducial cosmology: $H_0 \in \{65,66,67,67.7,69,70\}\;\hu$ and broader values: $H_0 \in \{40,50,60,67.7,80,90,100\}\;\hu$. For each catalog, we compare the detector-frame primary mass distribution and the merger rate $R(z)$. 
Both when slightly and strongly varying $H_0$, the detector-frame primary mass distribution remains compatible with the fiducial volume constrained by GWTC-4.0 analyses \cite{gwtc4pop, gwtc4cosmo}. This behavior can be seen in Fig.~\ref{fig H0 variation plots}(a) and in Fig.~\ref{fig H0 variation plots}(c). Thus, assuming a fixed, incorrect $H_0$ within the range of our interest does not push the population outside the relevant volume for real data, allowing to correctly perform spectral sirens analysis introduced in Sec.~\ref{HBI e NF}, without introducing biases.
The merger rate $R(z)$ shows an almost negligible dependence on $H_0$, as can be seen in Fig.~\ref{fig H0 variation plots}(b) and in Fig.~\ref{fig H0 variation plots}(d). This is expected for three different reasons:
\begin{enumerate}
   
\item  looking at Eq.~\ref{merger redshift eq}, we can distinguish between three different cases: when $\Delta t_D \ll t_{\rm LB}(z_f)$ the merger occurs shortly after the formation and the $H_0$ dependence of $z_M$ is negligible, when $\Delta t_D > t_{\rm LB}(z_f)$ the binary has not merged by today so it does not enter the catalog, finally only when $\Delta t_D \sim t_{\rm LB}(z_f)$ a change in $H_0$ shifts $z_M$. In the \bpop catalogs we considered, the number of binaries entering the latter case is negligible.

 \item merger redshift samples are drawn from the distribution in Eq.~\ref{merger redshift samples eq}: the $H_0^3$ scaling in $R(z)$ cancels against the same scaling in $dV_c/dz$, leaving just a residual $H_0$ dependence through the merger efficiency, which is subdominant in the isolated channel ($\lesssim 2\%$ for $H_0$ between $50 \hu $ and $100$ \hu \cite{papergiovanni}). 

\item if we consider the lookback time definition:
    \begin{equation}
        t_{\rm LB}(z) = \frac{1}{H_0} \int_0^z \frac{dz'}{(1+z')\sqrt{\Omega_m(1+z')^3+\Omega_\Lambda}} ,
    \end{equation}
   and we define an auxiliary function $f$:
   \begin{equation}
       f(y, \Omega_m) = \frac{\left(\sqrt{1-\Omega_m} - \sqrt{1-\Omega_m[1-(1+y)^3]}\right)^{2/3}} {1+y},
   \end{equation}
one can show that Eq.~\ref{merger redshift eq} can be written as:
\begin{equation}
    1+z_M = \frac{e^{-x \sqrt{1-\Omega_m}}}{f(z_F, \Omega_m)} \sqrt[3]{\frac{4(1-\Omega_m)}{[q(z_F,\Omega_m) - e^{-3x\sqrt{1-\Omega_m}}]^2}},
    \label{eq H0 vs zm}
\end{equation}
where $x=\Delta t_D \,H_0=\Delta t_D/t_H$, $t_H$ is the Hubble time and $q(z_F,\Omega_m)=\Omega_m/f(z_F, \Omega_m)^3$. Fig.~\ref{H0vszm}(a) shows the behavior of Eq.~\ref{eq H0 vs zm} as a function of $H_0$ for different values of $z_F$ and $\Delta t_D$. If $\Delta t_D \ll t_H$, then $x \ll 1$ and $e^{-x \sqrt{1-\Omega_m}} \sim 1$, which implies that the $H_0$ dependence is approximately negligible. A residual $H_0$ dependence becomes visible for bigger $\Delta t_D$ and it increases with $z_F$, it is however negligible for $z_F \lesssim 6$. Fig.~\ref{H0vszm}(b) shows the distribution of the $z_F$ samples in the \bpop catalog: $\sim 90\%$ of the events have $z_F<6$, thus the region in which the $H_0$ dependence becomes non-negligible contains only a small fraction of the population. 
\end{enumerate}

We conclude that the \bpop catalogs are, to a good approximation, cosmology-independent in the $H_0$ range of interest. Thus, the measurement of $H_0$ is not circular and does not originate from the cosmological choice in the \bpop catalogs.

\newpage
\bibliographystyle{ieeetr}
\bibliography{mybib}

\end{document}